\documentclass[article,twocolumn,floatfix]{revtex4}

\usepackage{graphicx}
\usepackage{color}
\usepackage{wasysym}

\newcommand{\ket}[1]{\left| #1 \right\rangle}

\newcommand{\proj}[1]{| #1\rangle\!\langle #1 |}

\newcommand{\Lm}{{\cal L}_{{\rm max}}}

\newcommand{\be}{\begin{equation}}
\newcommand{\ee}{\end{equation}}
\newcommand{\bea}{\begin{eqnarray}}
\newcommand{\eea}{\end{eqnarray}}

\begin{document}

\title{Error models in quantum computation: an application of model selection}
\author{Lucia Schwarz} 
\author{S.J. van Enk}
\affiliation{Department of Physics\\
Oregon Center for Optics\\University of Oregon, Eugene OR 97403}
\begin{abstract}
Threshold theorems for fault-tolerant quantum computing assume that errors are
of certain types. But how would one detect whether errors of the ``wrong'' type occur in one's experiment, especially if one does not even know what type of error to look for? The problem is that for many qubits a full state description is almost impossible to analyze due to the exponentially large state space, and a full process description requires even more resources.
As a result, one simply {\em cannot} detect all types of errors.
Here we show through a quantum state estimation example (on up to 25 qubits) how to attack this problem using {\em model selection}. We use, in particular, the Akaike Information Criterion. 
The example indicates that the number of measurements that one has to perform before noticing errors of the wrong type scales polynomially both with the number of qubits and with the error size.
\end{abstract}

\maketitle

\section{Introduction}

In order to develop a quantum computer we need to be able to coherently control and read out a system of many qubits. Verifying how a particular experimental implementation of a quantum computer actually performs
will be straightforward once we can run a computation in a fault tolerant manner: we just check whether the answer produced by the computation is correct or not. But before that time arrives we will need to employ other, less conclusive types of tests.

There are two types of generic tests that provide useful information about many-qubit systems: multi-partite entanglement verification tests \cite{guehne2009} and randomized benchmarking \cite{randb2008,randb2011}. However, the information gained is somewhat unspecific: In both cases one may detect that something is wrong, but one will not find out what exactly is wrong.
Unfortunately,  there is no efficient procedure to figure out what exactly is wrong, simply because we cannot efficiently simulate a generic multi-qubit quantum process on a classical computer. (For smaller systems quantum tomography can be used, and even there one has to be careful with systematic and other errors \cite{errors2012,errors2013,enk2013}.)

For fault tolerant quantum computation \cite{gott1998} one does need to know not just how large the error probabilities are, but also whether they are of the right type.
This is because threshold theorems \cite{aharonov1997,knill1998,kitaev1997,steane2003,fowler2012} need to make explicit use of error models. For example,  the calculation of the error threshold may be based on a  ``local stochastic'' error model (for an introduction, see \cite{gott2009}). Errors correlated over a long range may then be disastrous. One mechanism by which such long-range errors might arise is as follows.  A laser field's phase and intensity always fluctuate, but, of course, if those fluctuations are always sufficiently small, the errors they cause will be corrected for by quantum error correction. But what if the fluctuations, for just a brief time interval, are large? Then all qubits which happened to have been accessed during that time interval have a much larger probability of error. The problem we consider is how one could notice the presence of such errors.

While there is no systematic and efficient method to solve this problem completely, there is an efficient and well-tested  method: model selection \cite{claeskens1993,burnham2002}. This term refers to a well-developed field of (classical) statistics and inference where the aim is to rank different (statistical) models, each meant to describe some given process. 
In the present context model selection can be summarized as follows: 
We design a few-parameter model that describes our predictions of all the processes and errors that occur in our experiment ---it may have a few noise parameters with a clear physical meaning, for instance--- and compare it with a much larger (but still far from exhaustive) model that includes many (but not all) possible types of errors. As long as the large model contains a number of parameters that scales moderately with the number of qubits, then it still can be analyzed, even for a few dozen qubits. If that large model is ranked higher than our few-parameter model, we conclude that errors occurred that we did not expect.

We are going to discuss an illustrative example of this model selection procedure. We simulate a quantum state estimation experiment on $N$ qubits, which is modeled after an actual experiment performed on 14 ions in an ion trap in which a 14-qubit GHZ state of high fidelity was generated \cite{blatt14}.
We will vary $N$ up to 25 and assume the goal is to generate a perfect GHZ state.
We take a 3-parameter model (with three noise parameters describing three different noise processes) as our standard error model and then take a model with ${\cal O}(N^3)$ parameters as the much larger error model, which includes many types of errors, although, obviously, not all ${\cal O}(4^N)$ possible ones.
We assume the data are generated by a ``true'' state of the  form 
\[\rho_{{\rm true}}=(1-q)\rho_{{\rm s.e.}}+q\rho_{{\rm g.e.}},\]
with  the subscript s.e. referring to ``standard error model'' and ``g.e.'' to the more general error model.
We investigate then the following issues: First, does the model selection procedure recognize that
the standard error model is indeed correct (i.e., ranked higher than the large general error model) when $q=0$?  Second, in the case that $q\neq 0$, how many measurements does one need to take before one notices that there is in fact an error that lies outside the standard error model? The last question splits naturally into two subquestions, namely, how that number scales with the number of qubits and how it scales with $q$.

The model selection method we use here is based on the Akaike Information Criterion (AIC) \cite{akaike1998}. This method is widely used outside of physics, and by now has been applied on various occasions within quantum information theory as well \cite{usami2003,lougovski2009,yin2011,guctua2012,enk2013}. 
Most model selection criteria compare the goodness of fit of each model while penalizing the number of parameters, thus possibly favoring simpler models. The AIC in particular has a clear meaning since it is derived purely from the principles of information (see next section). It has been found to perform better than the related Bayesian Information Criterion \cite{burnham2002} in quantum state and entanglement estimation \cite{lougovski2009}.

\section{Preliminaries}
\subsection{Model selection and AIC}
Suppose we have taken data and now wish to model
the underlying process that generated the data.
Our data contains some amount of information about the underlying process, but also  statistical fluctuations. 
How can we determine whether a model is a good description of the underlying process rather than of the statistical fluctuations? In general, models with more parameters will be fitting the data better but are also more likely to fit to the fluctuations, and models with {\em too many} parameters are {\em over}fitting.
One method to find a compromise between under- and overfitting was proposed by Akaike \cite{akaike1998}. He derived an expression for the estimated Kullback-Leibler divergence between one's model and the true underlying process. The Kullback-Leibler  divergence is expressed in terms of two probability distributions for the data, the ``true distribution'' $\{p_i\}$, and the distribution generated by our model,  $\{s_i\}$, as follows:
\begin{equation}
KL(p||s) = \sum_i p_i \log \frac{p_i}{s_i}.
\end{equation}
This is a measure for the distance between the two probability distributions $\{p_i\}$ and $\{s_i\}$. It is also called the \emph{relative entropy} and can be understood as the information that is lost if the model $\{s_i\}$ is used instead of the ``real'' distribution $\{p_i\}$. 

Of course, we do not know the true underlying distribution,  but, nonetheless, the Kullback-Leibler divergence can be estimated, as was shown by Akaike, using the observed frequencies. Namely, up to a constant that is the same for all models, he found the divergence to approximately equal
\begin{equation}
AIC = -2 \Lm + 2K.
\end{equation}
Here $K$ is the number of parameters of the model,  and  $\Lm$ is its maximum log-likelihood,
\begin{equation}
\Lm= \max_{\{p_k\}} \sum_k f_{k} \log{p_{k}}, 
\end{equation}
with $f_k$ the number of times outcome $k$ was observed and $p_k$ the probability according to the model of obtaining outcome $k$. Model selection now consists of calculating the AIC for different candidate models, with the lowest score corresponding to the best model.
In our context this procedure can distinguish between models that accurately describe the relevant physical (error) processes, and models that spend too many parameters on fitting statistical noise. We thereby gain insights into the actual physical processes that cause errors, and we can tell whether or not errors outside our simple model are significant.

\subsection{A 3-parameter model for noisy GHZ states}

We will simulate an experiment on a noisy GHZ state \cite{greenberger1990} of $N$ qubits. The ideal GHZ state is a coherent superposition of all qubits in state $\ket{0}$ or all in state $\ket{1}$,
\begin{equation}
\ket{{\rm GHZ}} = \frac{1}{\sqrt{2}} \left(\ket{00...0}+\ket{11...1}\right). \label{ghz-state}
\end{equation}
A high fidelity version of this state was created in a trapped ion system for 14 ions \cite{blatt14}.
The density matrix $\rho_{{\rm GHZ}}=\proj{{\rm GHZ}}$, written in the standard basis $\ket{00...0}, \ket{00...1},...\ket{11...1}$, has only four nonzero elements  which all equal $\frac{1}{2}$. This state is maximally entangled and pure. However, a real quantum system in the lab will not be in this perfect state. There might be several effects that act on the qubits during the state preparation and/or storage. As a simple and not unreasonable model we assume just three noise processes, described by three parameters: a small imbalance $\varepsilon$ between the populations of $\ket{00..0}$ and $\ket{11..1}$, a systematic phase shift $\varphi$ of the relative phase between $\ket{00..0}$ and $\ket{11..1}$, and $\delta$ which quantifies the loss of coherence between
$\ket{00..0}$ and $\ket{11..1}$ due to random phase fluctuations. 
These three processes will create a mixed state with density matrix
\begin{equation}
\rho_{3P} =\frac{1}{2} \left( \begin{array}{ccc} 1+\varepsilon & ... & \delta \sqrt{1-\varepsilon^2} e^{i \varphi} \\
									\vdots & & \vdots \\
								  \delta \sqrt{1-\varepsilon^2} e^{-i \varphi} & ... & 1-\varepsilon 			 \end{array} \right).
\end{equation}

\subsection{A large model: permutationally invariant states}
For the comparison with our 3-parameter model, we try to find a model that will describe many possible errors and deviations from this simple model. However, we can't model \emph{all} possible errors that may occur. Our goal is therefore to design a model with a fairly large (but still polynomial) number of fitting parameters. If an arbitrary error is affecting our experiment, it will most likely be partially contained in this large model and we will detect it. Of course, this leaves out certain errors that are exactly orthogonal to the large model. To increase the chance of detecting small errors, for an actual experiment several such fairly large models should be considered and compared to each other. There is no systematic way to find good models for this purpose, but any model with a large number of parameters can be used. For simplicity, we only regard one such model in this paper. This suffices for our purpose of determining how many measurements are needed as a function of both $N$ and $q$.

The ideal GHZ state is permutationally invariant, in the sense that any permutation of its subsystems leaves the overall state unchanged. Mathematically, this can be expressed as
\begin{equation}
\rho_{{\rm GHZ}} = \frac{1}{N!} \sum_{\pi_k \in S_N} V(\pi_k) \rho_{{\rm GHZ}} V(\pi_k)^{\dagger},
\end{equation}
where the sum is over all $N!$ permutations $\pi_k$ of the $N$ qubits, and $V(\pi_k)$ is the unitary representation of the operator that permutes the subsystems according to the permutation $\pi_k$.
Since our simple 3-parameter model is also permutationally invariant, it makes sense to use as the large model the set of {\em all} permutationally invariant (PI) states. This set has been shown to be very convenient for quantum state reconstruction and entanglement detection \cite{toth2009,toth2010,moroder2012}. 
Many experiments aim at generating GHZ states, W states or Dicke states, all of which are PI. Moreover, if the PI part of a state $\rho$ is entangled, then so is $\rho$.

(Note that this choice does not imply that we think the actual state is permutationally invariant, and nor does it imply that we think the error process is permutationally invariant. All that matters is that our model will include the permutationally invariant part of the actual error process. 
As long as that part does not vanish, we will detect it. Recall that we {\em cannot} analyze all possible error models!)

As shown in \cite{moroder2012}, any permutationally invariant state can be represented as a block-diagonal matrix
\begin{equation}
\rho_{{\rm PI}} = \bigoplus_{j=j_{min}}^{N/2} P_j \rho_j \otimes \frac{\openone}{K_j},
\end{equation}
where $\rho_j$ is a spin-$j$ matrix of dimension $2j+1$ and $\{P_j\}$ is a probability distribution over the spin values $j$, and $K_j$ is the dimension of the non-PI part of the spin-$j$ state, given by
\be
K_j={N\choose N/2-j}-{N\choose N/2-j-1}.
\ee
The dimension of the permutationally invariant subspace grows as $\propto N^3$ with the number of qubits $N$. This model fits our purposes very well. For a dozen or more qubits the model contains a substantial number of parameters, but not so many that we cannot analyze it.

It was shown in Refs.~\cite{toth2010,moroder2012} that the necessary and sufficient number of different measurements needed to gain full information about a permutationally invariant state is
\be
D_N = { N+2 \choose N }.
\ee
In particular, we can choose to measure observables of the form $\hat{A}^{\otimes N}$, that is, we can measure the same single-qubit observable on each qubit. We just have to pick $D_N$ different single-qubit observables, the outcomes of which ought to be more or less uniformly distributed on the Bloch sphere \cite{toth2010,moroder2012}.

\section{Numerical results}
We simulate an experiment on $N$ qubits. The ``true'' state that generates the data is chosen to be an unequal mixture of a noisy GHZ state $\rho_{3P}$ (contained in the 3-parameter model) and a randomly picked permutationally invariant state $\rho_{{\rm PI}}$ orthogonal to the 3-parameter states (just to make sure the overlap of the actual state with the 3-parameter subspace does not vary with $N$). We write
\begin{equation}
\rho_{{\rm true}} = (1-q) \, \rho_{3P} + q \, \rho_{{\rm PI}}.
\end{equation}
The parameter $q$ determines the probability of ``wrong'' types of errors, namely, those outside our standard error model. 
\begin{figure}[h]
\includegraphics[width=\linewidth]{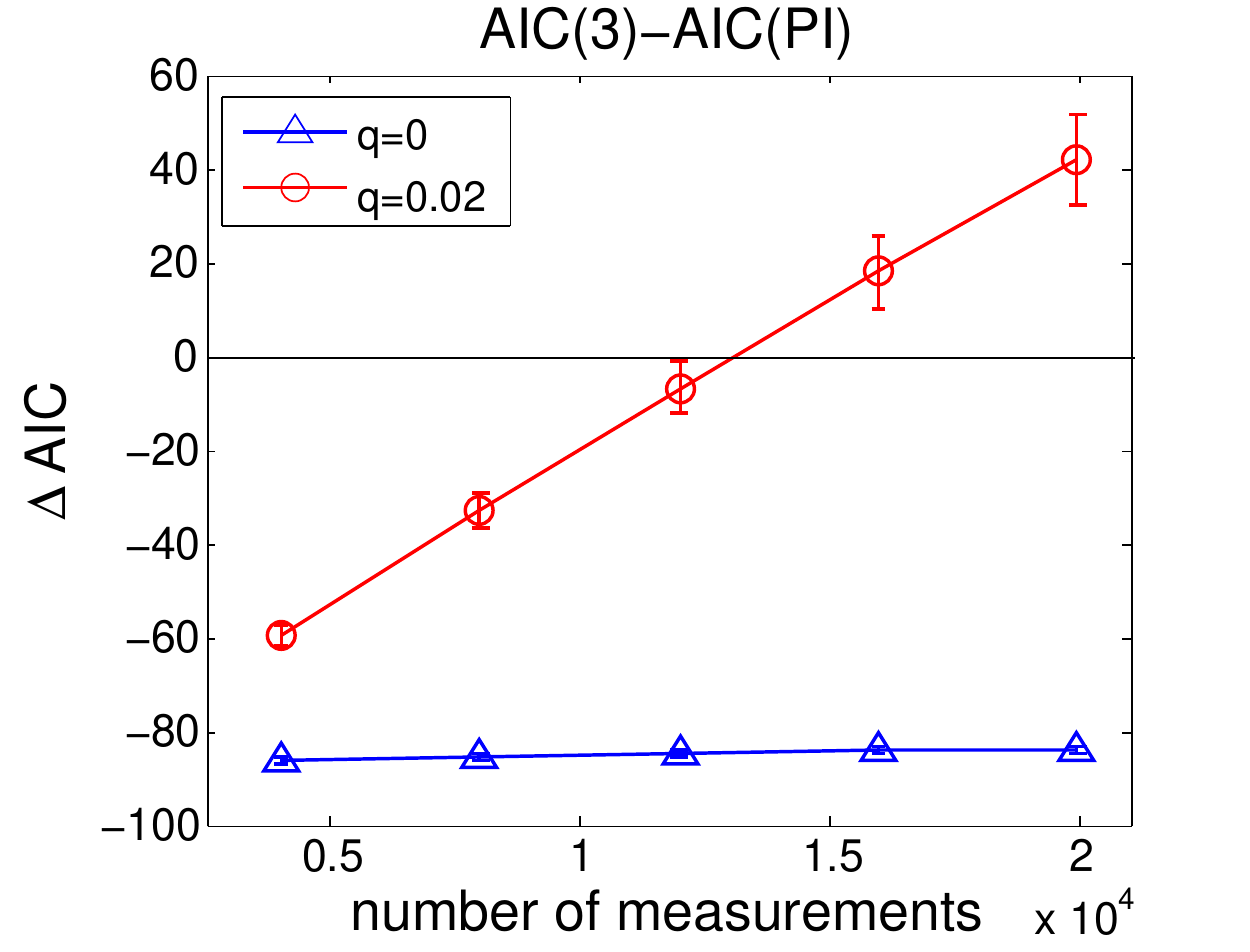}
\caption{(Color online) The differences in AIC values $\Delta AIC$ for a state of $N=5$ qubits plotted against the total number of measurements.  There are 21 measurements settings in this case, and the PI model contains 55 parameters. The simulation was run 100 times and the average $\Delta AIC$ is plotted. Error bars refer to the spread of $\Delta AIC$ over the 100 runs.}
\label{fig:aicM}
\end{figure}
We simulate a certain number of measurements,
where each single measurement consists of measuring $N$ times the same single-qubit observable, where the latter is chosen from the set of $D_N$ single-qubit observables. So a single measurement yields $N$ outcomes 0 or 1. We assume for simplicity that each of the $D_N$ observables is measured the same number of times.
We then find numerically the maximum likelihood state for the three-parameter model as well as for the large PI  model. This is easy for the three-parameter model since the minimization is over just $3$ parameters. For the PI model we apply an iterative algorithm described in \cite{moroder2012} for which the required computation time increases only polynomially in the number of qubits. Using the two maximum likelihoods thus obtained, we can calculate the respective AIC values for the 3P and PI models and plot the difference, which we denote by $\Delta AIC$. Negative values of $\Delta AIC$ correspond to the $3$-parameter model being favored, whereas positive values indicate that the PI model is better. 
\begin{figure}[h]
\includegraphics[width=\linewidth]{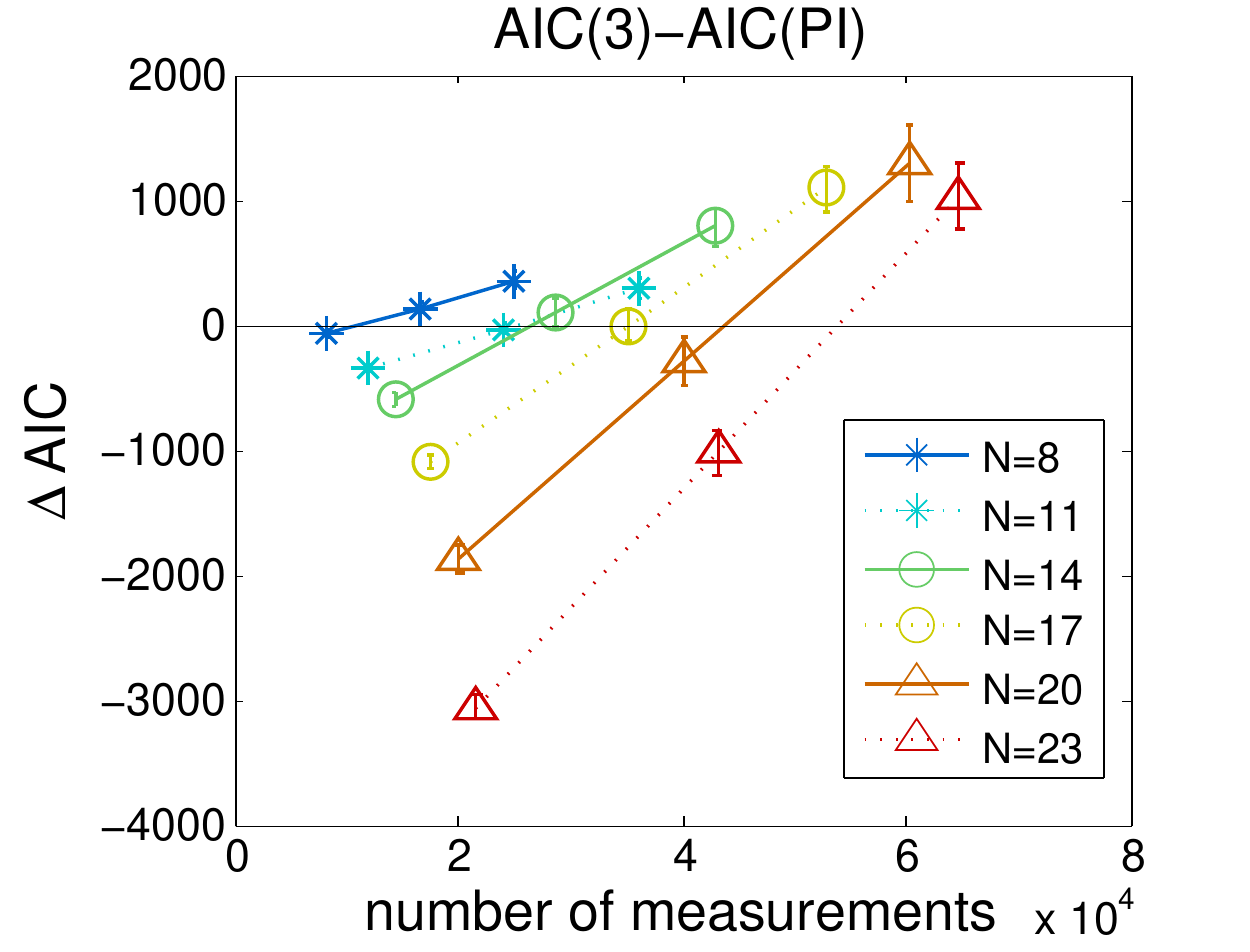}
\includegraphics[width=\linewidth]{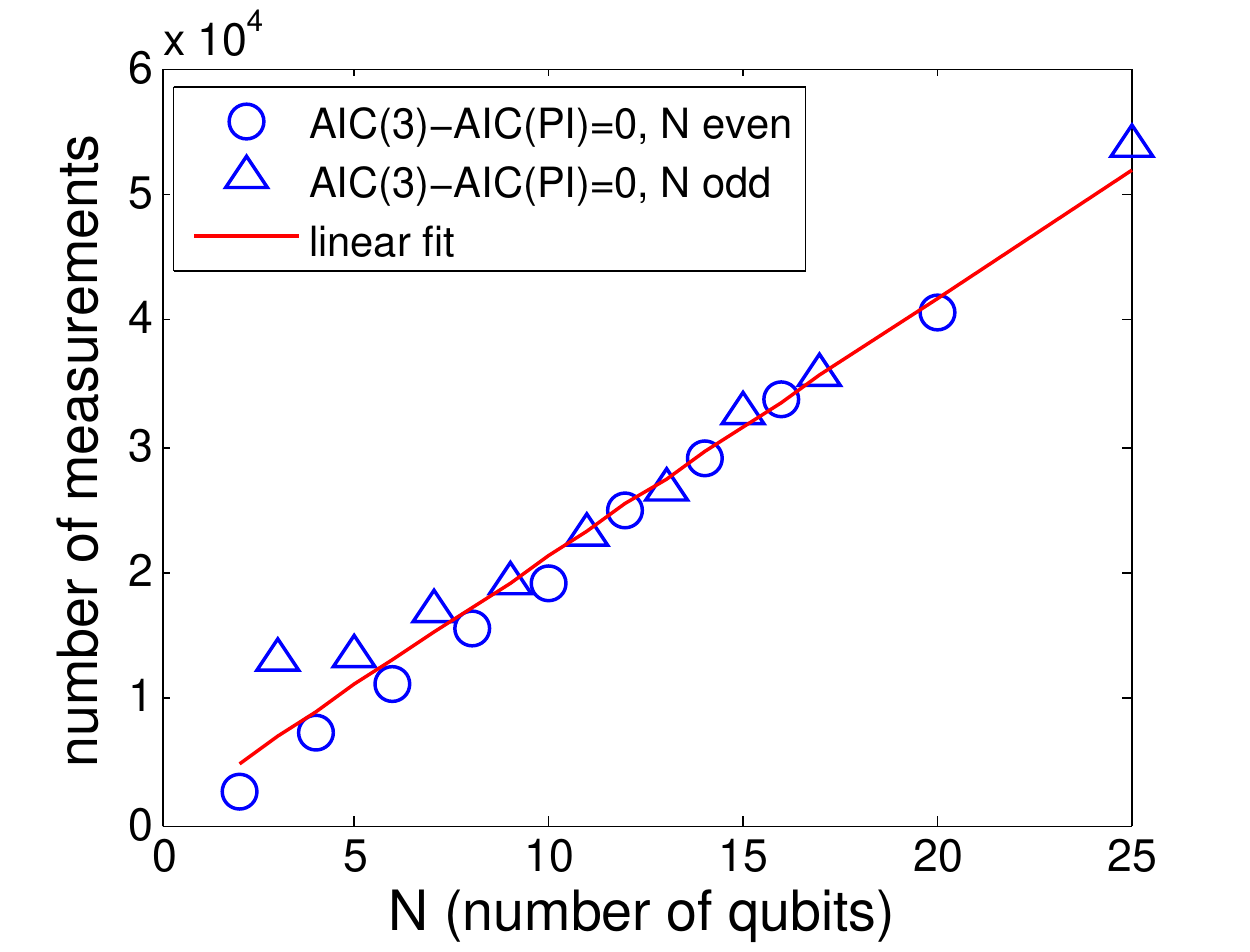}
\caption{(Color online) Plots for $q=0.02$: (a) The differences in AIC for several different numbers of qubits, plotted against the total number of measurements. Note that a single measurement on $N$ qubits yields $N$ binary outcomes. For very small numbers of measurements $\Delta AIC$ approaches twice the difference in the number of parameters of the two models  ($\approx N^3/3$). (b) The  average number of measurements required to reach the point where both models are rated equally. Small even and odd numbers of qubits behave slightly differently.}
\label{fig:aicN}
\end{figure}

Fig.~\ref{fig:aicM} shows $\Delta AIC$ for two different ``true'' states, one with $q=0$, the other with $q=0.02$. For $q=0$ the Akaike Information Criterion correctly always prefers the 3-parameter model. This is not as trivial (since the data are generated from a 3-parameter state!) as it may seem, because the statistical fluctuations are substantial (note that each observable is measured just a few dozen times for the smallest total number of measurements in the plot).
For $q=0.02$ we see that a relatively small number of measurements suffices to start favoring the PI model over the 3-parameter model, and the more measurements one performs, the firmer that conclusion gets. For very small numbers of measurements, a nonzero $q$ cannot be detected yet, and
we may interpret the point where $\Delta AIC$ crosses zero as the point where sufficiently many measurements have been taken to detect the presence of errors outside our standard (three-parameter) error model. 

Let us consider how that crossing point changes with the number of qubits.
A range of results for different $N$ is plotted in Fig.~\ref{fig:aicN}a. With increasing $N$ the crossing point clearly moves towards larger numbers of measurements.
We plot the crossing point as a function of $N$ 
in Fig.~\ref{fig:aicN}b.
We see that the necessary total number of measurements to detect a fixed perturbation $q$ increases only linearly in the number of qubits $N$, which shows that this can be measured very efficiently.
(The number of single-qubit measurements needed grows as $N^2$.)

\begin{figure}[h]
\includegraphics[width=\linewidth]{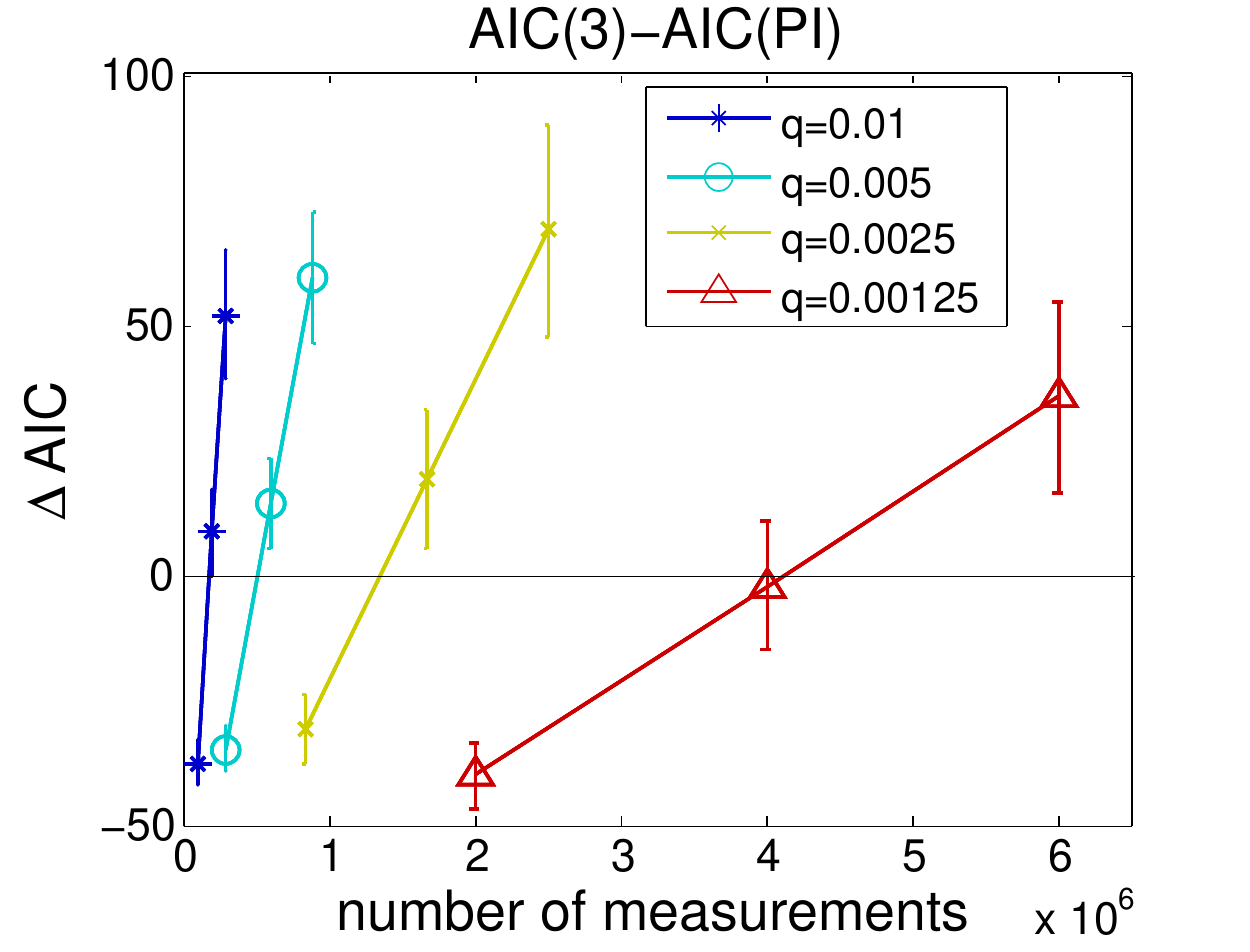} 
\includegraphics[width=\linewidth]{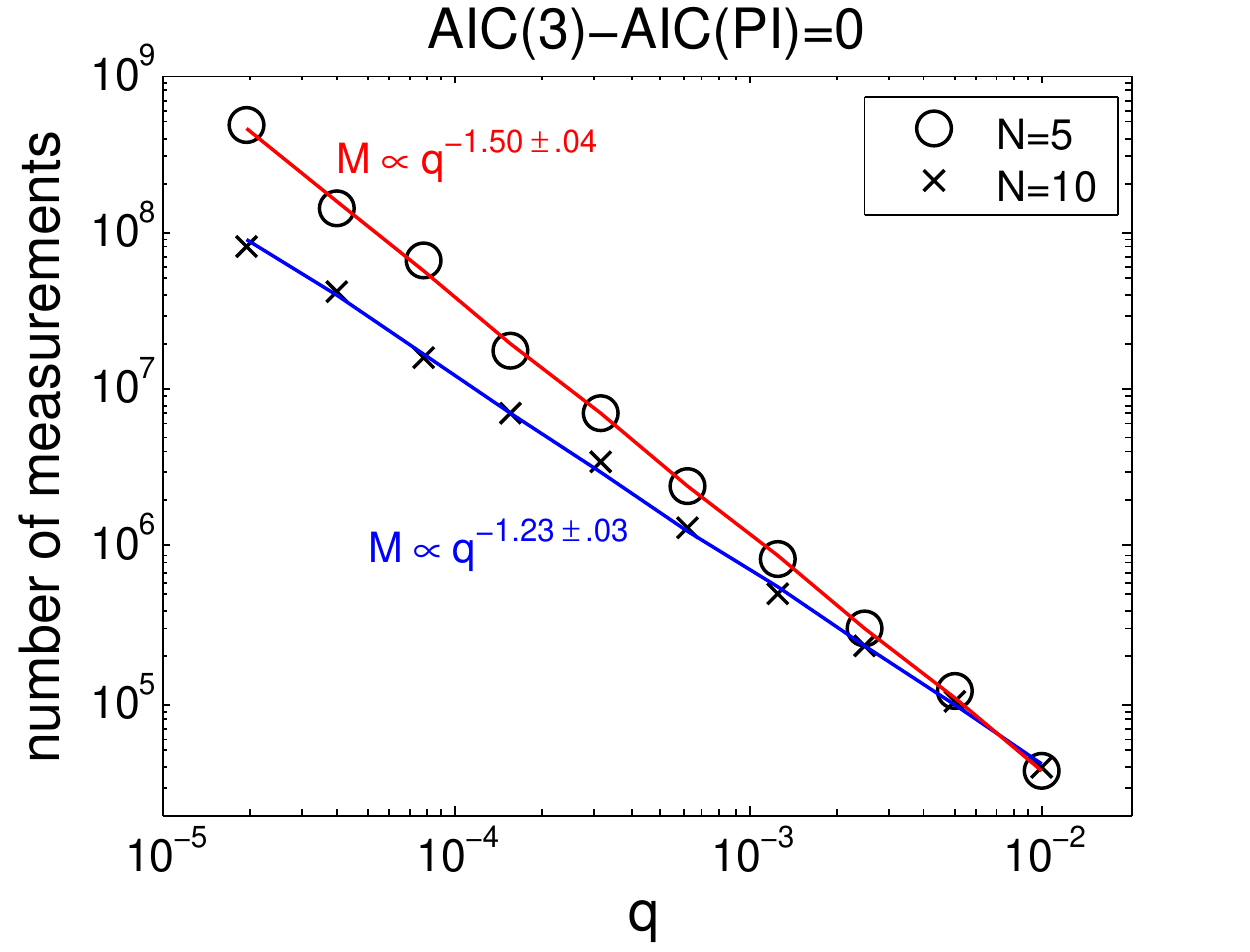} 
\caption{(Color online) (a) $\Delta AIC$ as a function of the number of measurements performed, for four different values of $q$ and $N=5$ qubits; (b) The minimum number of measurements $M$ needed to detect a perturbation of strength $q$, both for $N=5$ and for $N=10$ qubits. }
\label{fig:aicQ}
\end{figure}

It is also useful to investigate how the number of measurements needed to detect a nonzero value of $q$ depends on that value.

The plots of Fig.~\ref{fig:aicQ} show that the total number of measurements needed increases only moderately with $1/q$. This dependence becomes more favorable with increasing $N$, presumably because there are more ways to detect errors that occur with a given probability. 
\section{Conclusions}
We showed by example how to use the Akaike Information Criterion (AIC) to select between different error models in the context of quantum computing.  Thanks to the AIC one does not need exponentially many parameters to describe an experiment on multiple qubits. Instead, we compared a small model (with 3 parameters) with an intermediate-sized model (${\cal O}(N^3)$ parameters).
The former stands for a standard error model in the context of fault tolerant quantum computing, the larger model 
stands for other (undesired) types of errors. Since it is crucial to know whether one's implementation satisfies the condition for the fault tolerance error threshold theorems to apply, our method, which works for dozens of qubits, should be quite useful here.
In our specific example the number of (unentangled) $N$-qubit measurements needed to detect errors of the wrong type turned out to scale linearly with the number of qubits  and less than quadratically with the inverse of the wrong error probability. The latter scaling even improves with increasing number of qubits.
\section*{Acknowledgments}
We thank Austin Fowler for useful comments on an earlier draft of this paper.
This work was supported by the NSF Grant No. PHY-1004219.

\bibliography{aicpaper2}

\begin{thebibliography}{25}
\expandafter\ifx\csname natexlab\endcsname\relax\def\natexlab#1{#1}\fi
\expandafter\ifx\csname bibnamefont\endcsname\relax
  \def\bibnamefont#1{#1}\fi
\expandafter\ifx\csname bibfnamefont\endcsname\relax
  \def\bibfnamefont#1{#1}\fi
\expandafter\ifx\csname citenamefont\endcsname\relax
  \def\citenamefont#1{#1}\fi
\expandafter\ifx\csname url\endcsname\relax
  \def\url#1{\texttt{#1}}\fi
\expandafter\ifx\csname urlprefix\endcsname\relax\def\urlprefix{URL }\fi
\providecommand{\bibinfo}[2]{#2}
\providecommand{\eprint}[2][]{\url{#2}}

\bibitem[{\citenamefont{G{\"u}hne and T{\'o}th}(2009)}]{guehne2009}
\bibinfo{author}{\bibfnamefont{O.}~\bibnamefont{G{\"u}hne}} \bibnamefont{and}
  \bibinfo{author}{\bibfnamefont{G.}~\bibnamefont{T{\'o}th}},
  \bibinfo{journal}{Physics Reports} \textbf{\bibinfo{volume}{474}},
  \bibinfo{pages}{1} (\bibinfo{year}{2009}).

\bibitem[{\citenamefont{Knill et~al.}(2008)\citenamefont{Knill, Leibfried,
  Reichle, Britton, Blakestad, Jost, Langer, Ozeri, Seidelin, and
  Wineland}}]{randb2008}
\bibinfo{author}{\bibfnamefont{E.}~\bibnamefont{Knill}},
  \bibinfo{author}{\bibfnamefont{D.}~\bibnamefont{Leibfried}},
  \bibinfo{author}{\bibfnamefont{R.}~\bibnamefont{Reichle}},
  \bibinfo{author}{\bibfnamefont{J.}~\bibnamefont{Britton}},
  \bibinfo{author}{\bibfnamefont{R.}~\bibnamefont{Blakestad}},
  \bibinfo{author}{\bibfnamefont{J.}~\bibnamefont{Jost}},
  \bibinfo{author}{\bibfnamefont{C.}~\bibnamefont{Langer}},
  \bibinfo{author}{\bibfnamefont{R.}~\bibnamefont{Ozeri}},
  \bibinfo{author}{\bibfnamefont{S.}~\bibnamefont{Seidelin}}, \bibnamefont{and}
  \bibinfo{author}{\bibfnamefont{D.}~\bibnamefont{Wineland}},
  \bibinfo{journal}{Physical Review A} \textbf{\bibinfo{volume}{77}},
  \bibinfo{pages}{012307} (\bibinfo{year}{2008}).

\bibitem[{\citenamefont{Magesan et~al.}(2011)\citenamefont{Magesan, Gambetta,
  and Emerson}}]{randb2011}
\bibinfo{author}{\bibfnamefont{E.}~\bibnamefont{Magesan}},
  \bibinfo{author}{\bibfnamefont{J.}~\bibnamefont{Gambetta}}, \bibnamefont{and}
  \bibinfo{author}{\bibfnamefont{J.}~\bibnamefont{Emerson}},
  \bibinfo{journal}{Physical review letters} \textbf{\bibinfo{volume}{106}},
  \bibinfo{pages}{180504} (\bibinfo{year}{2011}).

\bibitem[{\citenamefont{Moroder
  et~al.}(2012{\natexlab{a}})\citenamefont{Moroder, Kleinmann, Schindler, Monz,
  G{\"u}hne, and Blatt}}]{errors2012}
\bibinfo{author}{\bibfnamefont{T.}~\bibnamefont{Moroder}},
  \bibinfo{author}{\bibfnamefont{M.}~\bibnamefont{Kleinmann}},
  \bibinfo{author}{\bibfnamefont{P.}~\bibnamefont{Schindler}},
  \bibinfo{author}{\bibfnamefont{T.}~\bibnamefont{Monz}},
  \bibinfo{author}{\bibfnamefont{O.}~\bibnamefont{G{\"u}hne}},
  \bibnamefont{and} \bibinfo{author}{\bibfnamefont{R.}~\bibnamefont{Blatt}},
  \bibinfo{journal}{arXiv preprint arXiv:1204.3644}
  (\bibinfo{year}{2012}{\natexlab{a}}).

\bibitem[{\citenamefont{Langford}(2013)}]{errors2013}
\bibinfo{author}{\bibfnamefont{N.~K.} \bibnamefont{Langford}},
  \bibinfo{journal}{New Journal of Physics} \textbf{\bibinfo{volume}{15}},
  \bibinfo{pages}{035003} (\bibinfo{year}{2013}).

\bibitem[{\citenamefont{van Enk and Blume-Kohout}(2013)}]{enk2013}
\bibinfo{author}{\bibfnamefont{S.~J.} \bibnamefont{van Enk}} \bibnamefont{and}
  \bibinfo{author}{\bibfnamefont{R.}~\bibnamefont{Blume-Kohout}},
  \bibinfo{journal}{New Journal of Physics} \textbf{\bibinfo{volume}{15}},
  \bibinfo{pages}{025024} (\bibinfo{year}{2013}).

\bibitem[{\citenamefont{Gottesman}(1998)}]{gott1998}
\bibinfo{author}{\bibfnamefont{D.}~\bibnamefont{Gottesman}},
  \bibinfo{journal}{Physical Review A} \textbf{\bibinfo{volume}{57}},
  \bibinfo{pages}{127} (\bibinfo{year}{1998}).

\bibitem[{\citenamefont{Aharonov and Ben-Or}(1997)}]{aharonov1997}
\bibinfo{author}{\bibfnamefont{D.}~\bibnamefont{Aharonov}} \bibnamefont{and}
  \bibinfo{author}{\bibfnamefont{M.}~\bibnamefont{Ben-Or}}, in
  \emph{\bibinfo{booktitle}{Proceedings of the twenty-ninth annual ACM
  symposium on Theory of computing}} (\bibinfo{organization}{ACM},
  \bibinfo{year}{1997}), pp. \bibinfo{pages}{176--188}.

\bibitem[{\citenamefont{Knill et~al.}(1998)\citenamefont{Knill, Laflamme, and
  Zurek}}]{knill1998}
\bibinfo{author}{\bibfnamefont{E.}~\bibnamefont{Knill}},
  \bibinfo{author}{\bibfnamefont{R.}~\bibnamefont{Laflamme}}, \bibnamefont{and}
  \bibinfo{author}{\bibfnamefont{W.~H.} \bibnamefont{Zurek}},
  \bibinfo{journal}{Science} \textbf{\bibinfo{volume}{279}},
  \bibinfo{pages}{342} (\bibinfo{year}{1998}).

\bibitem[{\citenamefont{Kitaev}(1997)}]{kitaev1997}
\bibinfo{author}{\bibfnamefont{A.~Y.} \bibnamefont{Kitaev}}, in
  \emph{\bibinfo{booktitle}{Quantum Communication, Computing, and Measurement}}
  (\bibinfo{publisher}{Springer}, \bibinfo{year}{1997}), pp.
  \bibinfo{pages}{181--188}.

\bibitem[{\citenamefont{Steane}(2003)}]{steane2003}
\bibinfo{author}{\bibfnamefont{A.~M.} \bibnamefont{Steane}},
  \bibinfo{journal}{Physical Review A} \textbf{\bibinfo{volume}{68}},
  \bibinfo{pages}{042322} (\bibinfo{year}{2003}).

\bibitem[{\citenamefont{Fowler}(2012)}]{fowler2012}
\bibinfo{author}{\bibfnamefont{A.~G.} \bibnamefont{Fowler}},
  \bibinfo{journal}{Phys. Rev. Lett.} \textbf{\bibinfo{volume}{109}},
  \bibinfo{pages}{180502} (\bibinfo{year}{2012}).

\bibitem[{\citenamefont{Gottesman}(2009)}]{gott2009}
\bibinfo{author}{\bibfnamefont{D.}~\bibnamefont{Gottesman}}, in
  \emph{\bibinfo{booktitle}{Quantum Information Science and Its Contributions
  to Mathematics, Proceedings of Symposia in Applied Mathematics}}
  (\bibinfo{year}{2009}), vol.~\bibinfo{volume}{68}, p.~\bibinfo{pages}{13}.

\bibitem[{\citenamefont{Claeskens and Hjort}(1993)}]{claeskens1993}
\bibinfo{author}{\bibfnamefont{G.}~\bibnamefont{Claeskens}} \bibnamefont{and}
  \bibinfo{author}{\bibfnamefont{N.~L.} \bibnamefont{Hjort}},
  \emph{\bibinfo{title}{Model selection and model averaging}}
  (\bibinfo{publisher}{Cambridge University Press}, \bibinfo{year}{1993}).

\bibitem[{\citenamefont{Burnham and Anderson}(2002)}]{burnham2002}
\bibinfo{author}{\bibfnamefont{K.~P.} \bibnamefont{Burnham}} \bibnamefont{and}
  \bibinfo{author}{\bibfnamefont{D.~R.} \bibnamefont{Anderson}},
  \emph{\bibinfo{title}{Model selection and multi-model inference: a practical
  information-theoretic approach}} (\bibinfo{publisher}{Springer Verlag},
  \bibinfo{year}{2002}).

\bibitem[{\citenamefont{Monz et~al.}(2011)\citenamefont{Monz, Schindler,
  Barreiro, Chwalla, Nigg, Coish, Harlander, H{\"a}nsel, Hennrich, and
  Blatt}}]{blatt14}
\bibinfo{author}{\bibfnamefont{T.}~\bibnamefont{Monz}},
  \bibinfo{author}{\bibfnamefont{P.}~\bibnamefont{Schindler}},
  \bibinfo{author}{\bibfnamefont{J.~T.} \bibnamefont{Barreiro}},
  \bibinfo{author}{\bibfnamefont{M.}~\bibnamefont{Chwalla}},
  \bibinfo{author}{\bibfnamefont{D.}~\bibnamefont{Nigg}},
  \bibinfo{author}{\bibfnamefont{W.~A.} \bibnamefont{Coish}},
  \bibinfo{author}{\bibfnamefont{M.}~\bibnamefont{Harlander}},
  \bibinfo{author}{\bibfnamefont{W.}~\bibnamefont{H{\"a}nsel}},
  \bibinfo{author}{\bibfnamefont{M.}~\bibnamefont{Hennrich}}, \bibnamefont{and}
  \bibinfo{author}{\bibfnamefont{R.}~\bibnamefont{Blatt}},
  \bibinfo{journal}{Physical Review Letters} \textbf{\bibinfo{volume}{106}},
  \bibinfo{pages}{130506} (\bibinfo{year}{2011}).

\bibitem[{\citenamefont{Akaike}(1998)}]{akaike1998}
\bibinfo{author}{\bibfnamefont{H.}~\bibnamefont{Akaike}}, in
  \emph{\bibinfo{booktitle}{Selected Papers of Hirotugu Akaike}}
  (\bibinfo{publisher}{Springer}, \bibinfo{year}{1998}), pp.
  \bibinfo{pages}{199--213}.

\bibitem[{\citenamefont{Usami et~al.}(2003)\citenamefont{Usami, Nambu, Tsuda,
  Matsumoto, and Nakamura}}]{usami2003}
\bibinfo{author}{\bibfnamefont{K.}~\bibnamefont{Usami}},
  \bibinfo{author}{\bibfnamefont{Y.}~\bibnamefont{Nambu}},
  \bibinfo{author}{\bibfnamefont{Y.}~\bibnamefont{Tsuda}},
  \bibinfo{author}{\bibfnamefont{K.}~\bibnamefont{Matsumoto}},
  \bibnamefont{and} \bibinfo{author}{\bibfnamefont{K.}~\bibnamefont{Nakamura}},
  \bibinfo{journal}{Physical Review A} \textbf{\bibinfo{volume}{68}},
  \bibinfo{pages}{022314} (\bibinfo{year}{2003}).

\bibitem[{\citenamefont{Lougovski and van Enk}(2009)}]{lougovski2009}
\bibinfo{author}{\bibfnamefont{P.}~\bibnamefont{Lougovski}} \bibnamefont{and}
  \bibinfo{author}{\bibfnamefont{S.~J.} \bibnamefont{van Enk}},
  \bibinfo{journal}{Physical Review A} \textbf{\bibinfo{volume}{80}},
  \bibinfo{pages}{052324} (\bibinfo{year}{2009}).

\bibitem[{\citenamefont{Yin and van Enk}(2011)}]{yin2011}
\bibinfo{author}{\bibfnamefont{J.}~\bibnamefont{Yin}} \bibnamefont{and}
  \bibinfo{author}{\bibfnamefont{S.~J.} \bibnamefont{van Enk}},
  \bibinfo{journal}{Physical Review A} \textbf{\bibinfo{volume}{83}},
  \bibinfo{pages}{062110} (\bibinfo{year}{2011}).

\bibitem[{\citenamefont{Gu{\c{t}}{\u{a}}
  et~al.}(2012)\citenamefont{Gu{\c{t}}{\u{a}}, Kypraios, and
  Dryden}}]{guctua2012}
\bibinfo{author}{\bibfnamefont{M.}~\bibnamefont{Gu{\c{t}}{\u{a}}}},
  \bibinfo{author}{\bibfnamefont{T.}~\bibnamefont{Kypraios}}, \bibnamefont{and}
  \bibinfo{author}{\bibfnamefont{I.}~\bibnamefont{Dryden}},
  \bibinfo{journal}{New Journal of Physics} \textbf{\bibinfo{volume}{14}},
  \bibinfo{pages}{105002} (\bibinfo{year}{2012}).

\bibitem[{\citenamefont{Greenberger et~al.}(1990)\citenamefont{Greenberger,
  Horne, Shimony, and Zeilinger}}]{greenberger1990}
\bibinfo{author}{\bibfnamefont{D.~M.} \bibnamefont{Greenberger}},
  \bibinfo{author}{\bibfnamefont{M.~A.} \bibnamefont{Horne}},
  \bibinfo{author}{\bibfnamefont{A.}~\bibnamefont{Shimony}}, \bibnamefont{and}
  \bibinfo{author}{\bibfnamefont{A.}~\bibnamefont{Zeilinger}},
  \bibinfo{journal}{American Journal of Physics} \textbf{\bibinfo{volume}{58}},
  \bibinfo{pages}{1131} (\bibinfo{year}{1990}).

\bibitem[{\citenamefont{T{\'o}th et~al.}(2009)\citenamefont{T{\'o}th,
  Wieczorek, Krischek, Kiesel, Michelberger, and Weinfurter}}]{toth2009}
\bibinfo{author}{\bibfnamefont{G.}~\bibnamefont{T{\'o}th}},
  \bibinfo{author}{\bibfnamefont{W.}~\bibnamefont{Wieczorek}},
  \bibinfo{author}{\bibfnamefont{R.}~\bibnamefont{Krischek}},
  \bibinfo{author}{\bibfnamefont{N.}~\bibnamefont{Kiesel}},
  \bibinfo{author}{\bibfnamefont{P.}~\bibnamefont{Michelberger}},
  \bibnamefont{and}
  \bibinfo{author}{\bibfnamefont{H.}~\bibnamefont{Weinfurter}},
  \bibinfo{journal}{New Journal of Physics} \textbf{\bibinfo{volume}{11}},
  \bibinfo{pages}{083002} (\bibinfo{year}{2009}).

\bibitem[{\citenamefont{Toth et~al.}(2010)\citenamefont{Toth, Wieczorek, Gross,
  Krischek, Schwemmer, and Weinfurter}}]{toth2010}
\bibinfo{author}{\bibfnamefont{G.}~\bibnamefont{Toth}},
  \bibinfo{author}{\bibfnamefont{W.}~\bibnamefont{Wieczorek}},
  \bibinfo{author}{\bibfnamefont{D.}~\bibnamefont{Gross}},
  \bibinfo{author}{\bibfnamefont{R.}~\bibnamefont{Krischek}},
  \bibinfo{author}{\bibfnamefont{C.}~\bibnamefont{Schwemmer}},
  \bibnamefont{and}
  \bibinfo{author}{\bibfnamefont{H.}~\bibnamefont{Weinfurter}},
  \bibinfo{journal}{Physical review letters} \textbf{\bibinfo{volume}{105}},
  \bibinfo{pages}{250403} (\bibinfo{year}{2010}).

\bibitem[{\citenamefont{Moroder
  et~al.}(2012{\natexlab{b}})\citenamefont{Moroder, Hyllus, Toth, Schwemmer,
  Niggebaum, Gaile, G{\"u}hne, and Weinfurter}}]{moroder2012}
\bibinfo{author}{\bibfnamefont{T.}~\bibnamefont{Moroder}},
  \bibinfo{author}{\bibfnamefont{P.}~\bibnamefont{Hyllus}},
  \bibinfo{author}{\bibfnamefont{G.}~\bibnamefont{Toth}},
  \bibinfo{author}{\bibfnamefont{C.}~\bibnamefont{Schwemmer}},
  \bibinfo{author}{\bibfnamefont{A.}~\bibnamefont{Niggebaum}},
  \bibinfo{author}{\bibfnamefont{S.}~\bibnamefont{Gaile}},
  \bibinfo{author}{\bibfnamefont{O.}~\bibnamefont{G{\"u}hne}},
  \bibnamefont{and}
  \bibinfo{author}{\bibfnamefont{H.}~\bibnamefont{Weinfurter}},
  \bibinfo{journal}{New Journal of Physics} \textbf{\bibinfo{volume}{14}},
  \bibinfo{pages}{105001} (\bibinfo{year}{2012}{\natexlab{b}}).

\end{thebibliography}

\end{document}